\documentclass[
  journal=largetwo,
  manuscript=article-type,
  year=2020,
  volume=37,
]{cup-journal}

\usepackage{upgreek}
\usepackage{aas-macros}
\usepackage{amsmath}
\usepackage[nopatch]{microtype}
\usepackage{booktabs}
\usepackage{gensymb}
\DeclareUnicodeCharacter{2212}{-}

\title{A Particle Detector Array deployed to the Murchison Widefield Array in the Murchison Radio-astronomy Observatory}

\author{J.E. Dickinson}
\affiliation{International Centre for Radio Astronomy Research, Curtin University, Bentley, 6102, WA, Australia}

\author{J.D.~Bray}
\affiliation{Jodrell Bank Centre for Astrophysics, Dept. of Physics \& Astronomy, University of Manchester, Manchester M13 9PL, UK}

\author{D.~Kenney}
\affiliation{International Centre for Radio Astronomy Research, Curtin University, Bentley, 6102, WA, Australia}

\author{T.~Booler}
\affiliation{International Centre for Radio Astronomy Research, Curtin University, Bentley, 6102, WA, Australia}

\author{J.~Edgley}
\affiliation{Jodrell Bank Centre for Astrophysics, Dept. of Physics \& Astronomy, University of Manchester, Manchester M13 9PL, UK}

\author{D.~Emrich}
\affiliation{International Centre for Radio Astronomy Research, Curtin University, Bentley, 6102, WA, Australia}

\author{A.~Forouzan}
\affiliation{International Centre for Radio Astronomy Research, Curtin University, Bentley, 6102, WA, Australia}

\author{T.~Gould}
\affiliation{International Centre for Radio Astronomy Research, Curtin University, Bentley, 6102, WA, Australia}

\author{A.~McPhail}
\affiliation{International Centre for Radio Astronomy Research, Curtin University, Bentley, 6102, WA, Australia}

\author{P.~Roberts}
\affiliation{Australia Telescope National Facility, CSIRO, Space and Astronomy, PO Box 76, Epping, NSW 1710, Australia}

\author{R.E.~Spencer}
\affiliation{Jodrell Bank Centre for Astrophysics, Dept. of Physics \& Astronomy, University of Manchester, Manchester M13 9PL, UK}

\author{L.~Verduyn}
\affiliation{International Centre for Radio Astronomy Research, Curtin University, Bentley, 6102, WA, Australia}

\author{R.~Watson}
\affiliation{Jodrell Bank Centre for Astrophysics, Dept. of Physics \& Astronomy, University of Manchester, Manchester M13 9PL, UK}

\author{A.~Williams}
\affiliation{International Centre for Radio Astronomy Research, Curtin University, Bentley, 6102, WA, Australia}

\author{K.~Grainge}
\affiliation{Jodrell Bank Centre for Astrophysics, Dept. of Physics \& Astronomy, University of Manchester, Manchester M13 9PL, UK}

\author{A.~Haungs}
\affiliation{Institute for Astroparticle Physics, Karlsruhe Institute of Technology, P.O. Box 3640, Karlsruhe, Germany}

\author{T.~Huege}
\affiliation{Institute for Astroparticle Physics, Karlsruhe Institute of Technology, P.O. Box 3640, Karlsruhe, Germany}
\alsoaffiliation{Astrophysical Institute, Vrĳe Universiteit Brussel, Pleinlaan 2, 1050 Brussel, Belgium}

\author{C.W.~James}
\affiliation{International Centre for Radio Astronomy Research, Curtin University, Bentley, 6102, WA, Australia}
\email[C.W.~James]{clancy.james@curtin.edu.au}

\author{S.J.~Tingay}
\affiliation{International Centre for Radio Astronomy Research, Curtin University, Bentley, 6102, WA, Australia}

\usepackage{biblatex}

\begin{document}

\begin{abstract}
We report the design and functionality of the Murchison Widefield Array Particle Detector Array (MWA PDA), an array of eight particle scintillation detectors deployed to Inyarrimanha Ilgari Bundara,
the Murchison Radio-astronomy Observatory (MRO). The purpose of the instrument is to identify cosmic ray extensive air showers (EAS) occurring over the core of the MWA radio telescope, and generate a trigger to allow radio data on the event to be captured and analysed. The system also acts as a pathfinder for a much larger instrument to be deployed in the core of the low-frequency component of the Square Kilometre Array, SKA-Low, by the SKA's ultra-high-energy particles science working group. Here, we describe the instrument and associated infrastructure, which has been verified to comply with the strict radio-frequency emissions requirements of the MRO, and was deployed in November 2024. We present calibration data, which demonstrates the ability of each detector to identify individual atmospheric muons at the expected rate, and we characterise the temperature dependence of the system. We describe a sample of 35,500 EAS identified using multi-detector coincidence over a 13-day period, and show how the detector data can be used to reconstruct the arrival directions and approximate energies of these events. We conclude that the PDA can reliably trigger on and reconstruct EAS contained within the $\sim 103 \times 90$\,m$^2$ core region, arriving within 20$^{\circ}$ of zenith, at primary cosmic ray energies above $\sim 4$\,PeV. We have also verified that the detector array can generate triggers, allowing the capture of radio data from the MWA correlator for offline analysis.
\end{abstract}

\section{Introduction}
Cosmic rays are highly energetic atomic nuclei that permeate interstellar space. These nuclei rain down on the Earth with the arrival directions of all but the most energetic homogenized by the Galactic magnetic field \citep[e.g.][]{IcetopAnisotropy,2017Sci...357.1266P}. Because the flux cannot be apportioned across identifiable sources, mechanisms of cosmic-ray production and propagation must be inferred from their energy spectrum and elemental composition.
Beyond approximately $10^{15}$\,eV the composition must be inferred using the properties of secondary particle air showers created by the interaction of the primary cosmic rays with the atoms of the atmosphere. Ground-based observations of a composition-dependent downturn in the cosmic-ray spectrum above the knee, i.e.\ in the $\sim$1--100\,PeV range \citep{KASCADEGrandeSpectrum}, reflect the rigidity dependence in leakage from the Milky Way \citep{CRLeakage}, or possibly time-dependence in nearby Galactic accelerators \citep{GalacticOrigin}, while a transition to a heavier composition at and above the ankle suggests a relatively low maximum injection energy for extragalactic sources \citep{AugerSpectrumComposition}.

Observations of the radio signature of cosmic ray extensive air showers have also been shown by the Low Frequency Array \citep[LOFAR;][]{schellart2013}, the Auger Engineering Radio Array \citep[AERA][]{2024PhRvL.132b1001A}, and others to provide precise measurements of shower development \citep{LOFARXmaxMethod2014}, and, hence, cosmic ray composition \citep{buitink2016}. The radio signal from EAS is due to currents induced by the Earth's magnetic field (the `geomagnetic effect'), and the rapid build-up of excess charge from electrons knocked into the cascade (the `Askaryan effect') --- see \citet{HuegeReview,FrankReview} for reviews. The emission is coherent in the forward direction, spread in a circle of a few hundred metres diameter projected onto the ground, and peaks below $100$\,MHz, so that precision studies require a dense array of low-frequency antennas.

In the near future, the Square Kilometre Array low-frequency telescope, SKA-Low, promises a massive increase in precision of radio measurements, allowing studies not just of cosmic ray composition, but, through tomographic studies of cascade development, disambiguation between hadronic interaction physics and composition models \citep{2017EPJWC.13502003H}.

The Murchison Widefield Array (MWA) is a precursor instrument for SKA-Low \citep{MWA_phaseI}, located in Inyarrimanha Ilgari Bundara, the Murchison Radio-astronomy Observatory (MRO). The MWA consists of 256 tiles with 16 antennas per tile, which can be beam-formed to a field-of-view (FOV) of 610\,deg$^2$ at 150\,MHz \citep{MWA_phaseII_design}. The full frequency range is 80--300\,MHz, of which 30.72\,MHz can be returned for processing. The characteristics that make the MWA well-suited to detecting cosmic ray radio emission are its dense core of 50 tiles within a 100\,m diameter --- complemented by two ``hexes'' of 72 tiles total offset $\sim$100--200\,m from the core --- and its large real-time buffering ability, which can store channelised tile voltage data for up to $\sim$2 minutes prior to ingestion by the correlator \citep{MWAX}.

Cosmic ray detection at the MWA serves as both a technical and scientific pathfinder for studies with SKA-Low. Technically, it aims to demonstrate how a particle detector array can be built with minimal radio-frequency emissions to satisfy the strict radio-quiet requirements of the Murchison Radio-astronomy Observatory (MRO) housing both MWA and SKA-Low, and operated in the harsh desert environment. Scientifically, it has the ability to probe higher frequency ranges than covered by LOFAR, where the polarisation signature of EAS has been verified below 100\,MHz \citep{LOFARPol} but is predicted to undergo a frequency-dependent change at higher frequencies to contain synchrotron-like components \citep{CoREAS,2022PhRvD.105b3014J,2024PhRvL.132w1001C}.

In order to study cosmic ray events, a method to synthesise the original signal from the 1.28\,MHz output channels of the MWA's polyphase filterbank has been developed \citep{2021JAI....1050003W}. An initial study using the MWA's voltage capture system on 26\,hr of data identified nine candidate cosmic ray events \citep{AlexThesis,2022icrc.confE.325W}. However, the extremely high raw data rate resulted in at most one hour's worth of data being recorded and analysed every few days, while the discrimination of cosmic ray events from RFI using radio-only data proved difficult. Therefore, an array of particle detectors --- the MWA particle detector array (MWA PDA) has been designed, built, and deployed to the MWA core, in order to detect EAS, and trigger the recording of MWA data whenever one occurs. The system is designed to be analogous to the LOFAR Radboud air-shower Array (LORA) detector used to trigger the LOFAR radio telescope \citep{LORA}.

In this work, we describe the eight particle detectors of the MWA PDA and their supporting infrastructure, which were deployed to the core of the MWA in November 2024. \S\,\ref{sec:PDA} describes the analogue components of the system, being the detectors themselves, their power supply, and cabling, all of which are installed near the core of the MWA. \S\,\ref{sec:DAQ} describes the data-acquisition system that identifies events in real-time, and which is installed at the central MWA Control Building.  \S\,\ref{sec:single} describes our calibration of the single-particle detector response, \S\,\ref{sec:performance} describes the performance of the detection system as a whole, while \S\,\ref{sec:analysis} describes an initial sample of detected cosmic ray events and their reconstructed properties; \S\,\ref{sec:conclusion} gives our concluding remarks on cosmic ray detection at the MRO.

\section{The Particle Detector Array}
\label{sec:PDA}

\begin{figure*}[hbt!]
\centering
\includegraphics[width=0.8\linewidth]{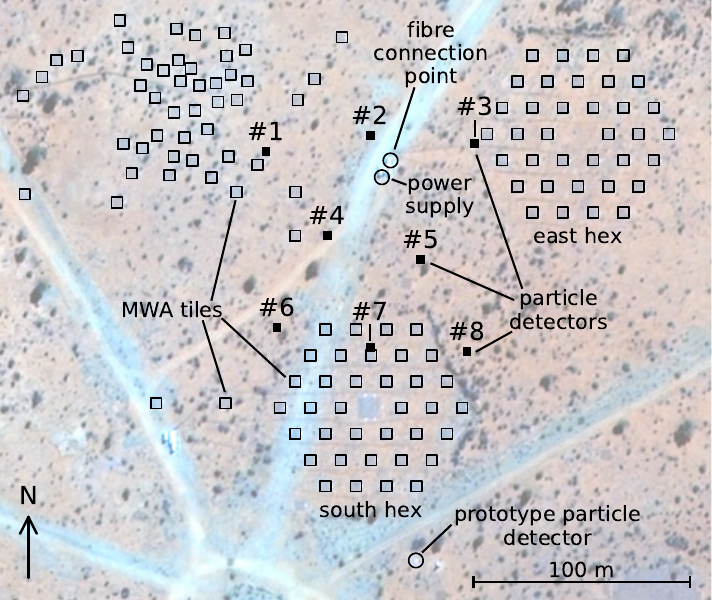}
\caption{An illustration of the distribution of the detectors \#1--8 of the PDA, situated between the tiles of the core and of the east and south hexes of the MWA.  All detectors are cabled to the power supply and fibre connection point shown.  Also marked is the prototype particle detector described in previous work \citep{Prototype}.  Background features in the image are access roads and vegetation.  Imagery from Siwei, processed by SkyFi.}
\label{fig:array}
\end{figure*}

The array deployed to trigger the acquisition of radio data associated with a cosmic ray air shower event comprises eight charged particle detectors distributed at approximately 50m intervals within the core of the MWA radio telescope \citep{tingay2013murchison, wayth2018phase}. The spacing was chosen to provide a similar level of coverage as the LOFAR Radboud Air-shower Array \citep[LORA; ][]{LORA}, which is used to trigger EAS detection at LOFAR, and to fit within the MWA core region. A total of eight detectors were specified to match the digitisation capabilities of our data acquisition system (DAQ) which was inherited from a previous experiment, and is described further in \S~\ref{sec:DAQ}.
Exact positions were constrained by cable lengths, terrain, and the positions of MWA tiles and other site infrastructure. Post-deployment, the positions were established using GPS to a precision of +/- 1m. The positions of the detectors are illustrated in Figure \ref{fig:array} and their latitudes and longitudes are presented in Table \ref{table:latlong}.

In addition to the particle detectors, the array infrastructure includes a power supply and field cabinet, also located within the core of the MWA, which powers the detectors; coaxial cables linked from the power supply to the detectors; and fibre-optic cable from the detectors to the MWA fibre-optic network.

The design of this system is broadly based on a single prototype detector, which is described in a previous work \citep{Prototype}.

\begin{table*}[hbt!]
\begin{threeparttable}
\caption{Data on the deployed detectors, giving detector positions in degrees decimal minutes (ddm), optical lengths of fibre-optic cable to the DAQ, corresponding time-delays relative to Detector 4 due to signal propagation along the fibre, and fibre link loss. The detectors are located at approximately 377.8\,m elevation above sea level.}
\label{table:latlong}
\begin{tabular}{lcccrc}
\toprule
\headrow  & Latitude  & Longitude & Fibre length & \multicolumn{1}{c}{$\Delta t$} & Loss\\
          & [ddm]       &    [ddm] & [m] & \multicolumn{1}{c}{[ns]} & [dB] \\
\midrule
Detector 1 &26° 42.068' S &116° 40.226' E & 5947.33 & 6.27 & 2.763\\ 
\midrule
Detector 2 &26° 42.064' S &116° 40.255' E & 5946.69 & 3.13 & 2.868\\ 
\midrule
Detector 3 &26° 42.066' S &116° 40.284' E & 5904.56 & -203.10 & 2.852\\ 
\midrule
Detector 4 &26° 42.089' S &116° 40.243' E & 5946.00 & 0.00 & 3.218\\ 
\midrule
Detector 5 &26° 42.095' S &116° 40.269' E & 5905.83 & $-$196.88 & 2.663\\ 
\midrule
Detector 6 &26° 42.112' S &116° 40.229' E & 5946.05 & 0.27 & 2.543\\ 
\midrule
Detector 7 &26° 42.117' S &116° 40.255' E & 5945.42 & $-$3.08 & 3.504\\ 
\midrule
Detector 8 &26° 42.118' S &116° 40.282' E & 5945.42 & $-$3.08 & 3.132\\ 
\bottomrule
\end{tabular}
\end{threeparttable}
\end{table*}

\subsection{The particle detectors}

\begin{figure}[hbt!]
\centering
\includegraphics[width=\columnwidth,clip=True,trim={80 0 80 0}]{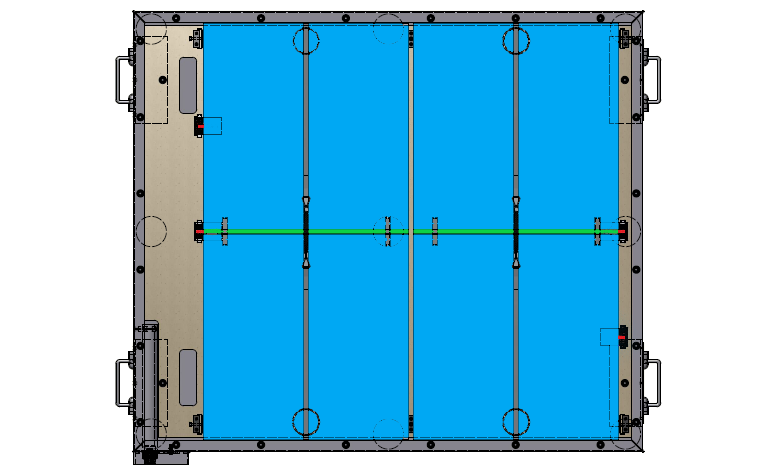}
\caption{An illustration of the interior of a particle detector. The scintillator material is coloured blue, the bars of wavelength shifter green, and the positions of the SiPMs red. The exit point for the fibre waveguide, and gland plate for supply voltage, are located in the lower left. The RFoF emitters sit in the space to the left of the scintillator. The nine larger circles are feet located on the underside of the box; the four smaller circles on the scintillators are plastic supports keeping the scintillator in place, with the aid of plastic straps (thick grey vertical lines) kept taut with springs (thick black lines crossing the green wavelength shifter). Four handles are attached for ease of transport.}
\label{fig:detector}
\end{figure}

The particle detectors contain slabs of Bicron BC416 plastic scintillator material and bars of Bicron BC482A wavelength shifter which were previously used in an identical arrangement in the trigger layer of the KASCADE central detector \citep{1997NuPhS..52...92K}. See Figure~\ref{fig:detector} for an illustration of the detectors.

Bicron BC416 has a density of 1.023\,g\,cm\textsuperscript{-3} and a refractive index of 1.58.\footnote{\url{https://luxiumsolutions.com/sites/default/files/2021-12/Organics-Plastic-Scintillators.pdf}} The data sheet for this material lists properties including a light decay time of 4\,ns and a pulse width of 5.3\,ns full-width half-maximum (FWHM). The wavelength of maximum emission is 434\,nm. Four square slabs of scintillator (47.5 x 47.5 x 3\,cm) are arranged in a square. These are treated as two pairs, with each pair separately covered in aluminium foil to reflect light back into the scintillator and to exclude stray light. The light yield of BC416 is $\sim$10,000 photons per MeV of energy deposited. A minimum ionizing muon loses energy at $\sim$2\,MeV cm\textsuperscript{2}\,g\textsuperscript{-1} and so would lose $\sim$6000\,keV when traversing the scintillator vertically,  yielding $\sim$60,000 photons. The bulk attenuation length of BC416 is specified as 400\,cm (1/e photons left) which is large when compared with direct photon paths to the photo-detectors, justifying the reflective coating on the slabs of scintillator in increasing the number of photons impinging on the photo-detectors.

The adjacent edges of the two pieces of each pair of scintillator sandwich a bar of wavelength shifter of size 47.5 x 3 x 1\,cm. The interface is shimmed to maintain a small gap between the scintillator and the wavelength shifter to facilitate total internal reflection of light emitted within the wavelength shifter. The wavelength shifter material is Bicron BC482A.\footnotemark[\value{footnote}] It has peak absorption at 420nm, well-matched to the peak in emission of the slabs of scintillator. It emits at 494nm with a decay time of 12ns and has a refractive index of 1.58. \citet{soler1993optical} report the absorption coefficient of BC480, a similar wavelength shifting material, to be $\sim$10 cm\textsuperscript{-1} around the wavelength of peak absorption, i.e.\ an attenuation length of $\sim$1mm. If this is also true for BC482A, then light of wavelengths around 420\,nm will be almost completely absorbed within the 1\,cm thickness of the wavelength shifter bar. The bulk attenuation length (for emitted photons) is again 400\,cm, long in comparison with the bar of wavelength shifter.

The photo-detectors are silicon photomultiplier (SiPM) arrays. Four SiPM arrays are installed in each detector. Two view the ends of the two bars of wavelength shifter that are sandwiched between the pairs of scintillator blocks. The other two view one of each pair of blocks of scintillator through a window at the midpoint of one side. The SiPM arrays are 2 x 2 arrays (ARRAYJ−60035−4P−BGA) of Onsemi J-series silicon photomultipliers,\footnote{Datasheet available at \url{https://www.onsemi.com/download/data-sheet/pdf/arrayj-series-d.pdf}} reflow soldered to a custom printed circuit board (PCB). The sensitive area of the SiPM array is 12.64 x 12.64 mm\textsuperscript{2}. The glass windows of the SiPMs have a refractive index of 1.53 at 436\,nm, which is well-matched to the refractive index of the scintillator and wavelength shifter. This would allow for efficient optical coupling using an appropriate index matching fluid, however the interfaces were left dry initially as the stability of such a fluid has not been established under the conditions that prevail on site.

Each of the four SiPMs of the array is tiled by 22,292 microcells, i.e.\ a total of 89,168 for each SiPM array. A microcell comprises a single photon avalanche diode (SPAD) which has the capacity to signal the absorption of a single photon. SPAD output is added in parallel in the SiPM, meaning that the output of a SiPM is proportional to the number of photons absorbed across the SiPM. In the SiPM array, the output of the four SiPMs is summed, increasing the sensitive area. 
An illustration of the arrangement of the slabs of scintillator, the bars of wavelength shifter, and the positions of the four SiPM arrays within a detector is shown in Figure~\ref{fig:detector}.

\begin{figure}[hbt!]
\centering
\includegraphics[width=\columnwidth]{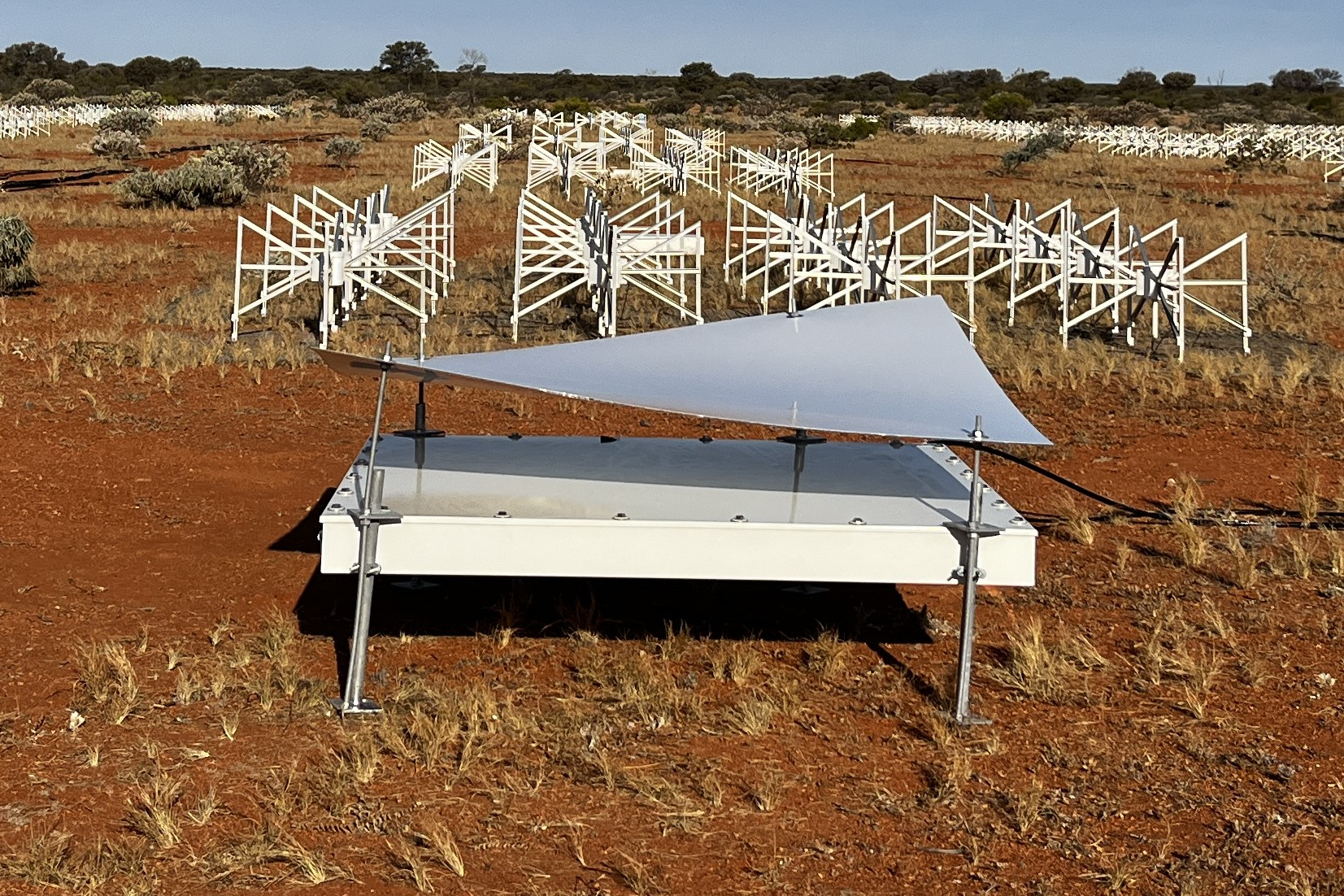}
\caption{A photograph of Detector 3 as deployed to the MWA.}
\label{fig:deployed}
\end{figure}

The active components of each particle detector are encased in a rectangular aluminium box that provides environmental protection and electromagnetic shielding. The cables supplying voltage to the detectors attach to a gland plate via SMA through connectors, and the signal exits on fibre via an aluminium waveguide. The detectors have legs which support them 20cm above ground to protect against flooding, and also a sun shade, shaped to reduce wind-induced vibrations, to ameliorate the effects of temperature variation in the electronic components. A photograph of one of the detectors deployed to the MRO is shown in Figure~\ref{fig:deployed}.

\subsection{Power supply and field cabinet}

An eight channel, custom-designed power supply provides power to the detectors. An Arduino microcontroller and associated digital to analogue converter (DAC) and ethernet boards, housed within a die-cast box inside the power supply, allow the voltage held on each channel to be specified in 0.01\,V intervals over a 12-bit range. A network connection is made with the Arduino over fibre via a media convertor and electrical network switch. The power supply, media converter, and network switch are all enclosed in a field cabinet that is designed to contain their electromagnetic emissions. An air-conditioner and networked power board are also housed within the field cabinet. The shielding performance of the cabinet, with all the components installed and working, was tested in an electromagnetic compatibility (EMC) chamber in Perth. The performance of the cabinet was assessed against the criterion that emission should be limited to 20\,dB below the MIL-STD461F RE102 “Navy Mobile \& Army” standard (MIL461-20) in accordance with the CSIRO policy \citep{7360651}. The output voltage for each detector is held across the cores of two coaxial cables within the field cabinet. Crystek CLPFL-0050 low-pass in-line filters were introduced at the gland plate of the field cabinet to prevent radio frequency noise being conducted out of the cabinet over the power cables.  

The power cables each consist of a twisted pair of 32 strands of 0.2\,mm diameter copper wire (i.e., a cross-sectional area of 1\,mm$^2$) 320\,m in length. Given the electrical resistivity of copper of $1.724\,10^{-8}\,\Omega$\,m at $20^{\circ}$\,C, and the 640\,m electrical path length for current return, this produces a resistance of $11.04\,\Omega$, i.e. a voltage drop of $\sim 1$\,V for the approximate current draw of 0.1\,A of each detector. Each cable is insulated with polypropylene, screened with aluminium/Mylar tape, sheathed in UV-stabilised polyethylene, and encased in a termite-resistant PVC jacket. This ensures survivability --- the cables are rated to 80$^{\circ}$\,C --- and prevents re-radiation of noise from the power-supply. Each cable is connected to a detector with coaxial cable tails.

The PDA power supply holds a voltage across the cores of two SMA bulkhead connectors on each detector, and this provides the power to a master PCB. This master-board supplies power to the four SiPM array PCBs in each detector. Each SiPM board supports a voltage across a SiPM array and provides signal amplification via a MAR-6SM+ inverting amplifier. Signal power attenuation of 2dB is also applied on the SiPM board.

\section{The Data Acquisition System (DAQ)}
\label{sec:DAQ}

The PDA data acquisition system (DAQ) is housed within the Correlator Room of the MWA Control Building. It comprises three components: a fibre receiver board, the Bedlam data processing board, and the bcc0a control computer.

\subsection{Signal transmission and the fibre receiver board}

The master-board in the detectors incorporates an ASTRON radio frequency over fibre (RFoF) optical transmitter module which converts the electrical signal from the SiPMs to optical for transmission from the field to the DAQ. The fibre receiver board converts the optical signal from the field back to electrical using a complementary optical receiver module. The distance over which the optical signal travels from each detector to the fibre receiver board was inferred using an optical time domain reflectometer (OTDR). Distances over fibre reported by the OTDR, and time length differences relative to signals from Detector 4 (for a refractive index of 1.4675 for the G.652.D step index fibre and 1310nm wavelength of the laser diode in the optical transmitters), are given in Table~\ref{table:latlong}. The losses measured in the fibre link by the OTDR (also reported in Table~\ref{table:latlong}) were approximately 3\,dB. The electrical path lengths from the fibre receivers to the Bedlam board are identical and the loss is assumed to be negligible.

\subsection{Bedlam data processing board}

\begin{figure}
    \centering
    \includegraphics[width=\linewidth,clip=True,trim={18cm 2cm 16cm 2cm}]{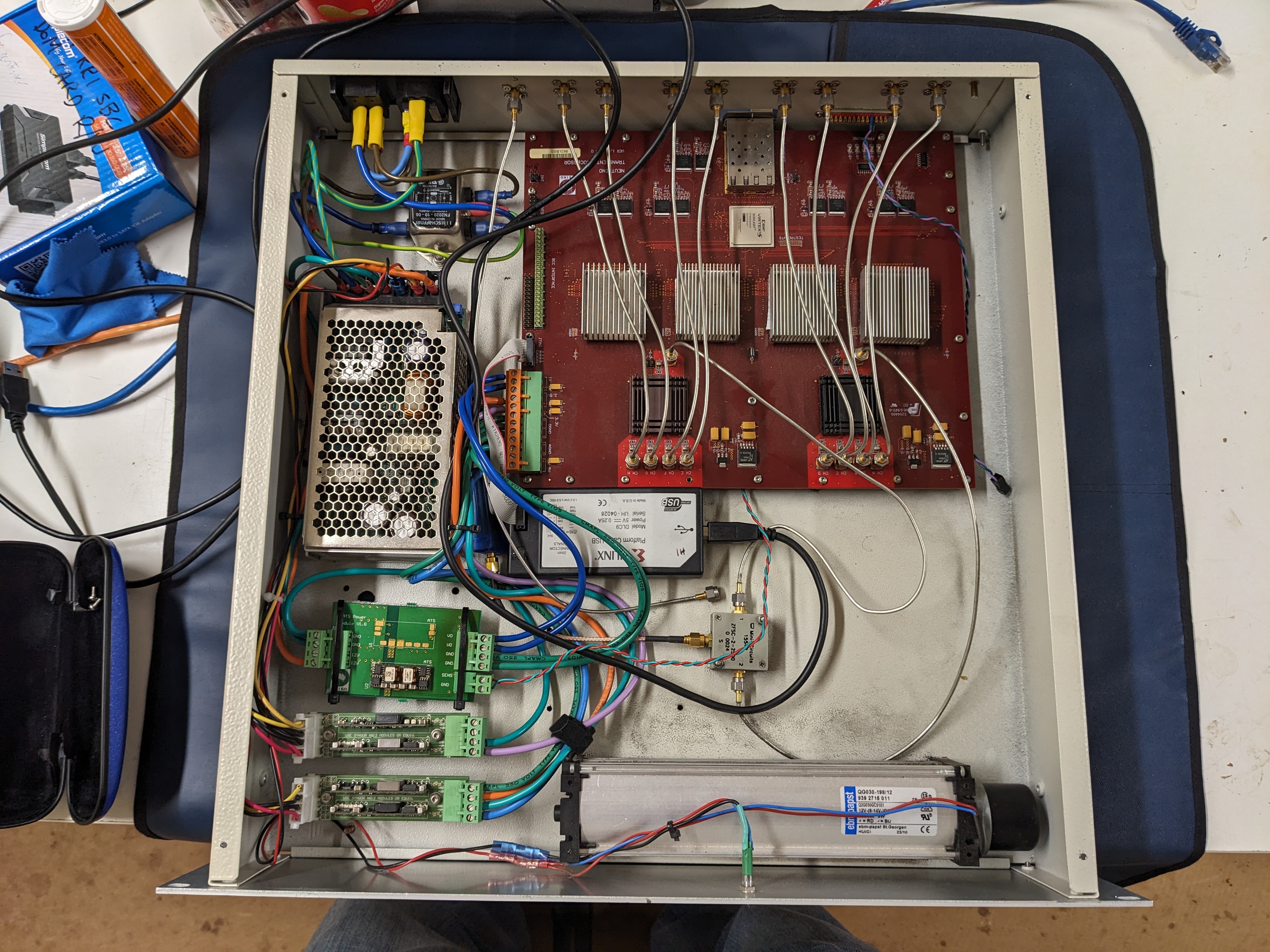}
    \caption{Photograph of the Bedlam digital signal processing board (top right) mounted in its rack while undergoing lab testing.}
    \label{fig:bedlam}
\end{figure}

The core of our DAQ is the Bedlam digital signal processing board \citep{Bedlam}, shown in Figure~\ref{fig:bedlam}. Developed for the LUNASKA series of experiments searching for ultra-high-energy particle interactions in the lunar regolith using the Parkes radio telescope \citep{2015APh....65...22B}, the Bedlam board was designed to perform digitisation and fast coincidence logic between eight analogue electrical inputs, with the signal in all eight channels being read only when (anti-)coincidence conditions were satisfied. The eight channels feed into two 8-bit, 4-channel analogue to digital converters (ADCs) and five Xilinx Virtex-5 family Field Programmable Gate Arrays (FPGAs). Four of the FPGAs dedicated to digital signal processing (model XC5VSX95T) receive the digitised data from the ADCs and apply an ADC threshold criterion to each. Data are digitised at 1.024\,GHz, with each ADC unit equating to 2mV of signal amplitude. The fifth FPGA is a logic orientated device (XC5VLX30T) that applies a coincidence requirement across the eight signal streams. The coincidence logic amounts to the requirement for a specified number of channels (i.e., between one and eight) to achieve their ADC thresholds within a specified time window. The signal received at the Bedlam board after passing through the optical link is bipolar. The Bedlam board has the capacity to apply a matched filter to the signal trace producing a predominantly positive--going pulse and suppressing noise. The digital signal processing FPGAs then apply the ADC thresholds to the absolute values of the filtered signal. A typical coincidence requirement for registering an event might be for three of the eight detectors to exceed an ADC threshold of 15 within a 4 micro-second window. The Bedlam board also records the rates that individual channels exceed their ADC thresholds, which can be used for control and monitoring purposes to set equal thresholds between detectors.

\subsection{Bedlam control computer (bcc0a)}

The Bedlam control computer (bcc0a) serves as the interface to Bedlam. It acts to set individual channel thresholds, multi-channel trigger requirements, and to monitor and adjust these on slow-control timescales of typically a minute. It also records the filtered and unfiltered signal traces for a specified time window --- typically 8192 samples, or $\sim 8.4\,\upmu$s --- whenever a coincidence is identified. A data file is opened for each minute within which an event occurs and all events that occur within that minute are recorded within that file. The Unix time at which the event occurred is also recorded. The time reference is the Network Time Protocol (NTP) time of the network that supports communication with the Bedlam control computer. This time reference is derived from a local CSIRO clock.

We have implemented and tested the ability of bcc0a to trigger the capture of radio data from the MWA using dummy triggers, with the MWA trigger to capture radio data being sent upon the receipt of a PDA trigger message from Bedlam. However, the goal of this present work is to finalise the criteria required for such a trigger, and we have not yet triggered and recorded radio data from a cosmic ray event from the MWA \citep[though such events have been detected in blind searches ][]{2022icrc.confE.325W}. When this mode is activated, radio data will be recorded and accessed through the MWA's All-Sky Virtual Observatory (MWA ASVO).\footnote{https://asvo.mwatelescope.org/}

The deadtime associated with data readout from Bedlam to bcc0a, during which further triggers cannot be recorded, was found during our prototyping phase to be at most 52.5\,ms for a maximum readout buffer size of 8192 samples \citep{Prototype}, which is negligible compared to the trigger rates considered in this work.

\section{Characterisation of SiPM response}
\label{sec:single}

Identification of an extensive air shower requires distinguishing it from the background of individual muons from low-energy cosmic rays, and from thermally generated noise from the SiPMs themselves. In this section, we analyse the behaviour of this noise, and the response of the SiPMs to individual muons.

\subsection{Dark counts}

\begin{figure}[hbtb]
\centering
\includegraphics[width=\columnwidth]{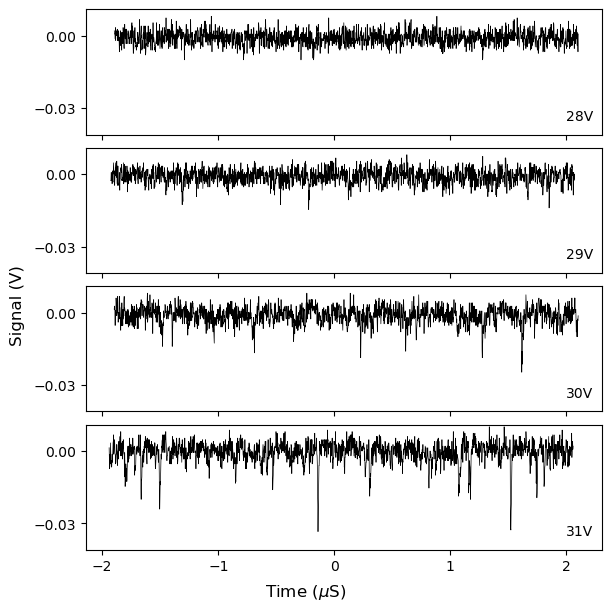}
\caption{Signals seen from a SiPM board in a dark box for four applied voltages. }
\label{fig:dark}
\end{figure}

The SiPM arrays within the detectors are exposed to large diurnal fluctuations in temperature which leave them susceptible to variation in the thermal noise in the SPADs. A thermally generated free charge carrier creates an avalanche in the SPAD that cannot be distinguished from an avalanche generated by a photon. The so-called ``dark noise count'' across a SiPM is an exponential function of temperature, and is primarily responsible for the $\sim30$\% diurnal fluctuations in count rate seen in our prototype detector. Fortunately, SPAD gain is a weak inverse function of temperature, and so an operating voltage for the SiPM arrays might be set at which the amplitude of single avalanches is lower than typical excursions in the signal due to instrumental noise. Events due to chance satisfaction of the coincidence requirement would then be dominated by instrumental noise.
To investigate the dark counts, a SiPM board was sealed in a dark box and the signal fed directly to a digital oscilloscope. Four-microsecond signal samples were taken, at a 1\,GHz sample rate, with four different voltages applied across the SiPM array. Typical examples of the signal observed over these periods are shown in Figure \ref{fig:dark}.

In Figure \ref{fig:dark}, dark events can be seen in the signal traces recorded for all voltages in excess of 28\,V applied at the SiPM board. At 28\,V, excursions from zero are more symmetrical about zero, demonstrating that the single thermal electron noise is masked by the noise intrinsic to the SiPM board, while at higher voltages, significant negative excursions corresponding to dark-count events become increasingly prevalent. A nominal operating voltage of 28\,V was therefore adopted for the SiPM arrays to eliminate the effects of dark counts on the PDA. We have yet to systematically test the operation of the array under different applied voltages.

The measurements were made in the lab at 28\degree C. At higher temperatures, the frequency of dark events will increase, but the gain of the SPADs of the SiPM arrays will decrease. Experience with our prototype detector showed the latter to be the dominant effect. At lower temperatures the converse will be true.

\subsection{Single particle response}

\begin{figure}
\centering
\includegraphics[width=1.0\linewidth]{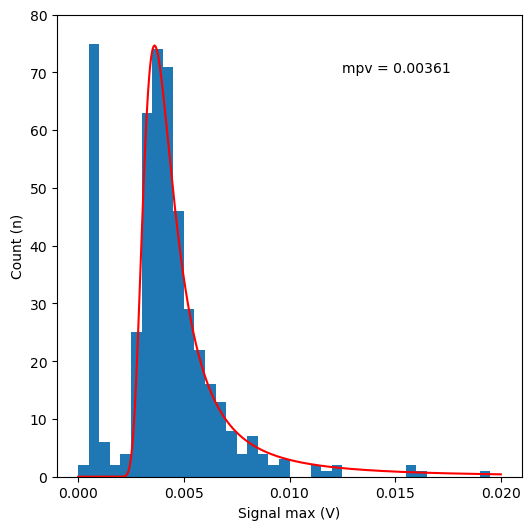}
\caption{An example histogram representing the distribution of peak signals from a SiPM board with its SiPM array viewing a bar of wavelength shifter sandwiched between two quadrants of scintillator within a detector. The data were collected on an oscilloscope using a muon paddle placed below one of the quadrants of scintillator as a trigger. The muon paddle was used to sample the sensitivity of the detector across the quadrant. Signal in the first few bins can be attributed to instrumental noise. The remaining data are fit by a Landau function (shown in red) to return the most probable value (mpv) for the response to a single charged particle.}
\label{fig:wavelength_shifter}
\end{figure}

An important consideration in the detector design is the response profile of the SiPM arrays to muons arriving across the slabs of scintillator. Our prototype detector --- which did not use wavelength-shifting bars --- was found to have a very inhomogenous response \citep[][; also verified using a muon tower at the Karlsruhe Institute of Technology]{Prototype}, with muons incident on the four corners of the scintillator being extremely difficult to detect. The response profile of our redesigned detectors was measured across a single quadrant of scintillator within a detector using a muon paddle placed below the detector as a trigger. The muon paddle comprised a 7.5 x 7.5 x 5.5\,cm block of BC408 scintillator housed in a dark box together with a SiPM board. The muon paddle was used to trigger a scope and was placed at 36 equally spaced positions under the detector to sample the response of the detector across the scintillator quadrant. The response of a SiPM board was recorded for 500 events at each point with 28\,V applied across the SiPM array. This was repeated for SiPM boards viewing the wavelength shifter, and those viewing the scintillator directly.

Figure\,\ref{fig:wavelength_shifter} shows an example histogram of the responses of a SiPM board with the SiPM array viewing the wavelength shifter sandwiched between two slabs of scintillator. The first few bins of each histogram represent instrumental noise, with the charged particle having missed the scintillator before arriving at the paddle. There is a clear minimum between the instrumental noise and the signal distribution demonstrating that the distribution can be attributed to the signal from individual particles. The first 4 bins (of width 0.0005\,V) were therefore not used to fit the signal distribution. A Landau distribution is then fit to the remaining data returning the most probable value (mpv) in the distribution. The mpv is stable across the scintillator with average mpv being 0.00358\,V with a standard deviation of 0.00012\,V, i.e.\ the response of the SiPM boards viewing the wavelength shifter can be considered independent of muon position within the scintillator slab. Since the wavelength shifter can receive photons from both scintillator slabs that sandwich it, and signals from the SiPM arrays viewing the two bars of wavelength shifter are summed by the master board, a detector is, therefore, equally sensitive to a single charged particle over its whole sensitive area.

\begin{figure}
\centering
\includegraphics[width=1.0\linewidth]{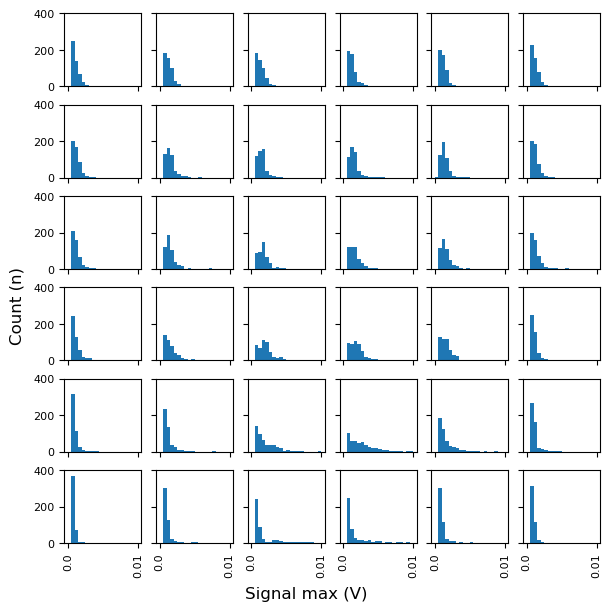}
\caption{Histograms representing the distributions of peak signal from a SiPM board whose SiPM array was viewing a quadrant of scintillator directly. The SiPM array was positioned at the bottom of the figure between the third and fourth columns of histograms.}
\label{fig:scintillator}
\end{figure}

Figure \ref{fig:scintillator} shows histograms of responses from a SiPM board whose SiPM array was viewing a quadrant of scintillator directly. The SiPM array was located along the bottom edge of the scintillator, i.e.\ at the bottom of the matrix of histograms in the figure, between columns 3 and 4. The distribution of events that contain signal due to light from the scintillator cannot be distinguished from events solely containing instrumental noise. The distributions are much more heterogeneous than those recorded for the SiPM arrays viewing the bars of wavelength shifter. Greatest sensitivity to a charged particle is constrained to a sector with its apex at the position of the SiPM array.

The flat response profile across the scintillator, and discrete single particle response, for the SiPM array viewing the bar of wavelength shifter, makes that arrangement appropriate for sampling extensive air showers at low particle densities. The response of SiPM arrays viewing the scintillator directly would be flatter across the detector for particle densities which would result in multiple particles traversing the scintillator in a single shower. Also, the SiPM arrays viewing the scintillator directly receive light from only a single quadrant of scintillator while those viewing bars of wavelength shifter receive signal from two quadrants. The differences between the responses of these two configurations lend themselves to providing low gain and high gain responses, respectively, allowing the dynamic range of the array to be extended. The signals from the SiPM boards with SiPM arrays viewing the scintillator directly are delayed by 50ns with respect to those viewing the bars of wavelength shifter. An additional 10\,dB of attenuation is applied to the prompt signals from the SiPM arrays viewing the wavelength shifter, and 30\,dB of attenuation to the delayed signals from the SiPM arrays viewing the scintillator directly. The attenuation is added at the output of the SiPM boards using Mini-Circuits VAT-10A+ and VAT-30A+ in-line attenuators, respectively. The signals from the four SiPM boards are then combined on the master-board. A second MAR-6SM+ inverting amplifier on the master board provides further amplification and transforms the signal into a positive going pulse. The signal is converted into an optical signal on the master board and leaves the detector on optical fibre.

\subsection{Gain from the SiPM board to the input to the Data Acquisition System (DAQ)}

\begin{figure}
\centering
\includegraphics[width=\columnwidth,clip=True,trim={0.2cm 0 10cm 0}]{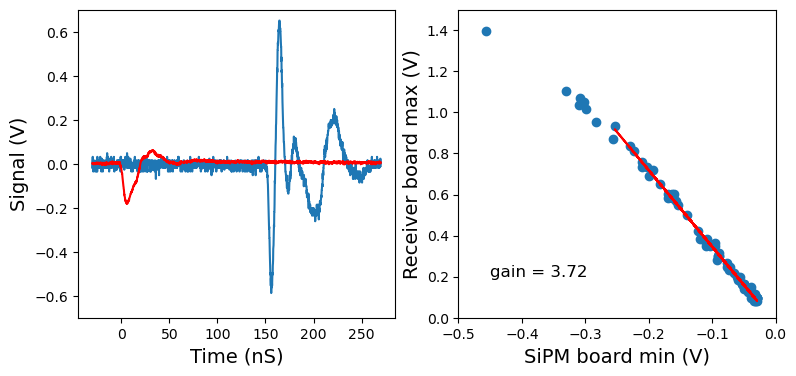}
\includegraphics[width=\columnwidth,clip=True,trim={10.1cm 0 0.1cm 0}]{Figures/Gain3.png}
\caption{The top panel illustrates the transformation of signal between the output of the SiPM board and the output of the optical receiver board. The signal at the output of the SiPM board is shown in red, and that at the output of the optical receiver board in blue. The baselines of both signals are centred on a voltage of zero. The signal path between these points contains a 10\,dB in-line attenuator, the master board which incorporates an optical transmitter, a short optical path, and the optical receiver board. The lower panel plots the maximum of the blue trace against the minimum of the red trace, taken over multiple signals, again relative to the baseline voltages at each location. The gain is given by the magnitude of the gradient of the linear fit (red line).}
\label{fig:gain}
\end{figure}

The mean of the most probable values for the Landau functions describing the response of the SiPM arrays viewing the wavelength shifter is 0.00358\,V. To allow extrapolation of the single particle response further through the signal chain, a measurement of the gain of the system from the output of a SiPM board to the output of the optical receiver board was made. A signal was taken from a muon paddle and split using a tee-connector at its output. The top panel of Figure \ref{fig:gain} shows an example event. The trace in red is taken directly from where the signal is split and that in blue from the output of the receiver board. The bottom panel shows the relationship between the largest negative value of the red trace and the largest positive value of the blue trace. The magnitude of the gradient of the line is the gain that can be attributed to the sum of the 10\,dB attenuator, the master-board, and the optical link. The gain, being a factor of 3.72 (5.7\,dB), is constant for signals up to 0.25V. The most probable value for the response to a single particle at the SiPM board (0.00358\,V) translates to an output at the receiver board of 0.0133\,V. An additional loss of $\sim$3\,dB was measured in the approximately 6\,km of fibre between the detectors and the control building. After accounting for this loss, the prompt signal for a single particle will equate to $\sim$5 ADC units in the Bedlam board, and the signal will saturate at slightly more than 25 particles. This measurement was made with the SiPM at 28\,V, the nominal operating voltage.

\subsection{Single detector particle response rate}

\begin{figure}
\centering
\includegraphics[width=1\linewidth]{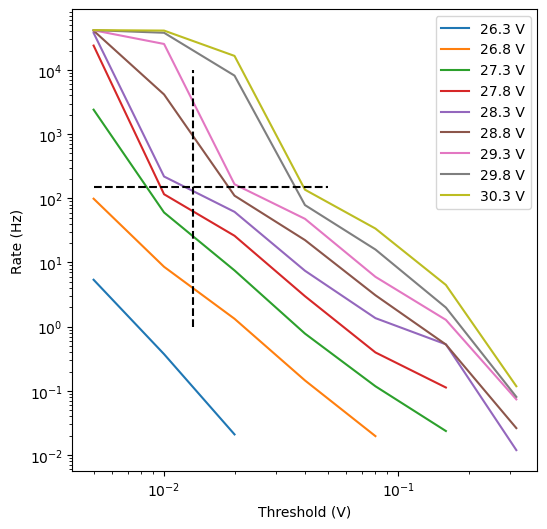}
\caption{The rates at which signals from a complete detector are exceeded for a range of thresholds and SiPM voltages. The vertical dashed black line represents the most probable signal amplitude (0.0133V) for a muon traversing the scintillator. The horizontal dashed black line shows the muon flux through the detector sensitive area of $\sim$0.9m$^{2}$ for a muon flux at sea-level of 1 muon per square cm per minute. The saturation of the rates at around 4 x 10\textsuperscript{4} is the maximum rate at which the oscilloscope could process the data segments.}
\label{fig:rate}
\end{figure}

A measurement of the rate at which the signal from a complete detector exceeds a range of threshold voltage amplitude values was made over a range of SiPM input voltages (the voltage supplied to the SiPM was verified at the receiver board). The optical output of the detector was passed over fibre to the fibre receiver board and the rates at which a particular thresholds were exceeded were measured on an oscilloscope. The results are displayed in Figure \ref{fig:rate}. While the rates generally decrease as a power-law with increasing threshold voltage, an inflection point can be seen at the approximate background atmospheric muon rate, where the rate transitions from being noise-dominated to muon-dominated. As expected, the known single muon rate is approximately recovered at the most probably value of the Landau function (translated to DAQ voltage, i.e.\ 0.0133\,V) when the SiPM voltage is set to $\sim$28.3\,V --- a slightly higher voltage is needed to recover muon signals below the most probable value. This informed our initial trigger settings for multi-particle coincidence to identify EAS in the core of the MRO.

\section{Performance}
\label{sec:performance}

The array of eight detectors was deployed during the period of 4$^{\rm th}$--8$^{\rm th}$ Nov.\ 2024. After a period of initial testing, the array ran continuously for a period of approximately 13 days in the same configuration from February $18^{\rm th}$ 08:04 UTC to March $2^{\rm nd}$ 12:44 UTC, i.e.\ a period of 12 days, four hours, and 40 minutes, at which point we deemed that sufficient data had been collected for a first-look analysis.
During this period, the coincidence requirement for an event to be recorded was for three of the eight detectors to exceed an ADC threshold in the filtered data buffer of 15 within a time window of 4\,$\upmu$s --- note that this ADC value is not calibrated against the voltages measured over specific components shown in Figure~\ref{fig:gain}. The power supply was set to maintain a nominal 28\,V across the SiPMs within each of the detectors. This resulted in a total of 36,868 events being recorded. 

\begin{figure}
\centering
\includegraphics[width=\columnwidth]{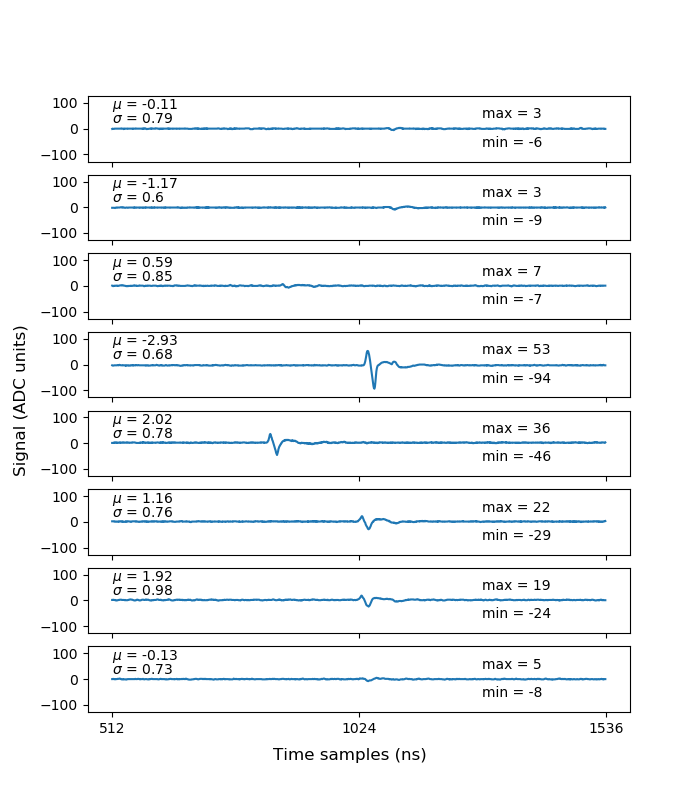}
\caption{Raw data describing an event recorded on the 18th February 2025 at 09:59 UTC. Signal is evident in all eight of the traces. The channels number off from top to bottom. The largest signals are seen in Detectors 4, 5, 6 and 7 and it is likely that the core of the shower was within the array. The mean, $\mu$, and standard deviation, $\sigma$, values are calculated from the first 512 samples in the signal buffer, and the maximum and minimum signal heights are also reported. The signals in Channels 1, 2, 3 and 8 are consistent with response to a single particle.}
\label{fig:Example}
\end{figure}

Figure \ref{fig:Example} shows the raw data buffers for a single event. Signal is evident in all of the channels, with the highest amplitudes --- corresponding to the core of the shower --- registering in Detectors 4 and 5. The signals in Channels 3 and 5 appear advanced in time relative to the rest of the array, as expected from their shorter fibre signal paths. Also shown on the figure are the means and standard deviations of the first 512 bins of the raw buffers.

\subsection{Means and standard deviations of system noise}

\begin{figure}
\centering
\includegraphics[width=\columnwidth]{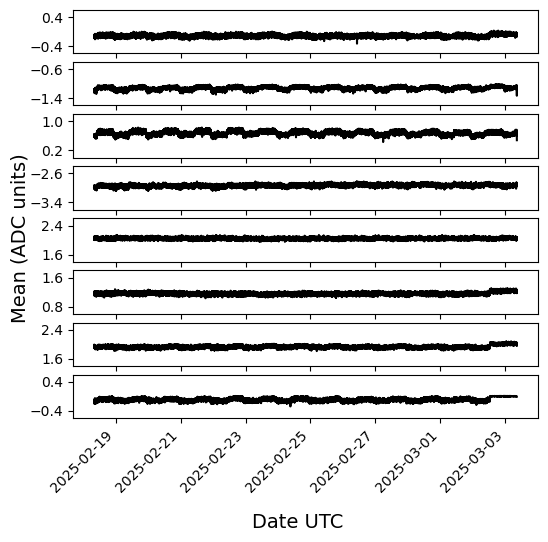}
\caption{Means of 512 bin samples of system noise. Means for Detectors 1-8 are shown in rows 1-8 respectively.}
\label{fig:Mean}
\end{figure}

Figure \ref{fig:Mean} shows the means of 512 samples of the signal noise recorded in each event across the recording period. The means are seen to be systematically displaced from zero, and some diurnal variation is seen in some channels. Consequently, the mean of 512 bins of system noise in each event and for each detector was removed from the signal prior to further analysis. Within Bedlam, DC offsets can be programmed and subtracted from the signal prior to triggering --- however, there is currently no mechanism to automatically monitor these offsets, so we only account for them in offline analysis.

\begin{figure}
\centering
\includegraphics[width=\columnwidth]{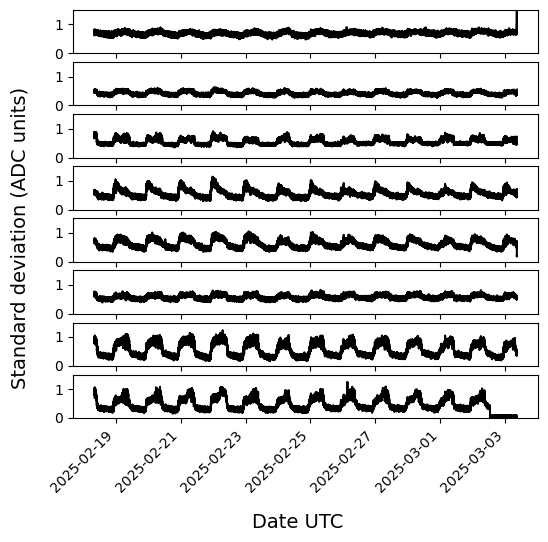}
\caption{Standard deviations of 512 bin samples of system noise}
\label{fig:standard_deviations}
\end{figure}

Figure \ref{fig:standard_deviations} shows the standard deviation of 512 bins of system noise from each event. A diurnal variation is also seen in these data. The standard deviation might be seen as a surrogate for gain, in which there is known diurnal variation --- certainly, a positive correlation between standard deviation and air temperature was seen in our prototype detector. 

Also evident in the mean and standard deviation data is the failure of Detector 8 towards the end of the observing period, due to a failure of a single channel of the power supply. The dataset under analysis was truncated to exclude events that were compromised. This reduced the number of events in the dataset to 35,500.

\begin{figure}
\centering
\includegraphics[width=1\linewidth]{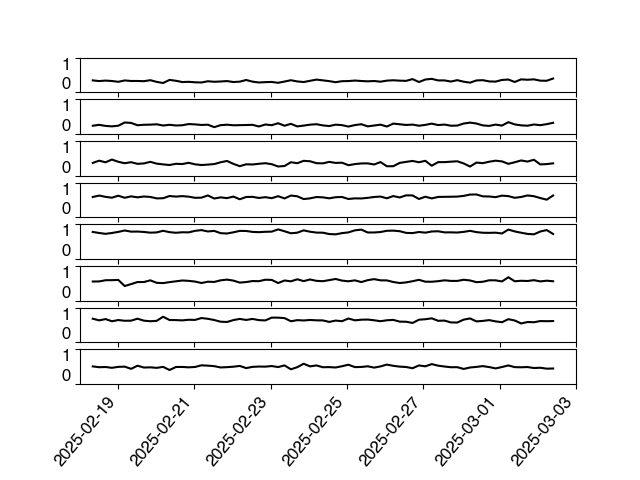}
\caption{Fraction of triggered events with a contributing signal in each detector (plotted inrows 1-8 respectively), calculated in four-hourly intervals.}
\label{fig:fractions}
\end{figure}

We also show in Figure~\ref{fig:fractions} that fraction of events in which each detector contributes to the trigger, calculated every four hours for the latter sample of events in which all eight detectors were working. The relative fractions are a combination of both individual detector properties, and the array geometry.

We do not attempt to explain differences in these behaviours between detectors, and attribute  them to the sum of minor differences in various detector components.

\begin{figure*}
\centering
\includegraphics[width=0.8\linewidth]{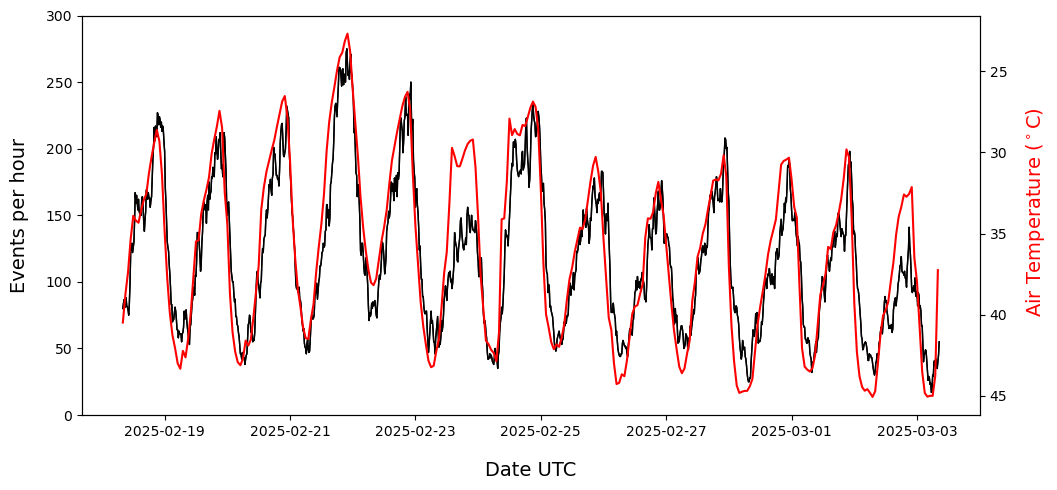}
\caption{The event rate across the 13 days that the array was operational is shown in black. The peaks in the event rates correspond to early morning at the MRO when temperatures were lower. The red trace shows the air temperature at the site on an inverted scale.}
\label{fig:Event_rate}
\end{figure*}

\subsection{Temperature sensitivity}

Figure \ref{fig:Event_rate} shows the event rate across the period that the array was operational. It is evident that there was a significant diurnal variation in the event rate. The same diurnal variation in the standard deviation in the system noise was previously identified --- and correlations with air temperature were also identified during our prototype experiments --- and so a measure of the on-site air temperature was obtained from the ingest air temperature of a component of the MWA. This varied between 25\,\degree C and 45\,\degree C over the duration of the deployment. It is also plotted in Figure \ref{fig:Event_rate}, inversely scaled to highlight the close linear correspondence between event rate and air temperature. The breakdown voltage of the p-n junction in the SiPMs increases at a rate of 21.5 mV/\,\degree C, leading to a corresponding decrease in the over-voltage across the junction. A 20\,\degree C increase in temperature can, therefore, result in a 0.43\,V reduction in over-voltage and concomitant reduction in gain. There are, however, other potential influences that might result in a decrease in signal with temperature, one being an increase in the resistance of the copper power cables supplying the detectors. Such effects can be addressed by adjusting the voltages across the SiPMs to stabilize the rates at which the detectors exceed their ADC thresholds, a performance indicator that can be read from the device registers of the Bedlam data processing board. We leave such adjustments to a future work.

\section{Cosmic ray event reconstruction}
\label{sec:analysis}

The analysis in previous sections suggests that the majority of the 35,500 events in the sample will represent detection of EAS. However, before using the PDA to trigger the MWA, we verify the fidelity of our measurements by demonstrating that PDA measurements are consistent with an EAS origin. This is particularly important when selecting events that lie within the MWA beam, and will thus have readily interpretable radio signals, or identifying events during periods of high RFI, where data from the PDA itself may be required to help identify the RF signature of EAS.

\subsection{Cosmic-ray arrival direction reconstruction}
\label{sec:arrival_direction}

The PDA is small in comparison with typical air shower detectors, and this limits its potential for analysing the shape of the shower front. In fact, in the configuration it operated, a large proportion of the events were only detected by three of the detectors. Consequently, a planar wavefront was assumed, and the normal to the wavefront calculated from the arrival times at the detectors on the ground, offset by the known fibre delays. The fit used a simple $\chi^2$ minimisation.

\begin{figure}
\centering
\includegraphics[width=\columnwidth]{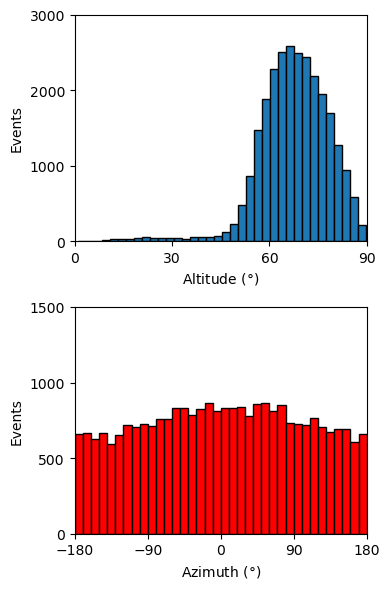}
\caption{Altitude (horizon at $0^{\circ}$, zenith at $90^{\circ}$) and azimuthal (North at $0^{\circ}$, East at $+90^{\circ}$) distributions of reconstructed event arrival directions.}
\label{fig:Alt_azi}
\end{figure}

The distributions of the altitude and azimuthal angles of the normal vector are shown in Figure \ref{fig:Alt_azi}. The distribution of azimuthal angle is close to uniform, with a preference for events arriving from the North ($0^\degree$) rather than South. Such a monopolar excess cannot be due to irregularities in the locations of the detectors or their sensitivities, since this would produce a dipolar preference for e.g.\ NS-elongated ground patterns over EW-elongation, and such elongation would be near-symmetrical with azimuthal angles separated by $180^\degree$. We suspect that a further timing error, which we are yet to identify, is responsible.

\subsection{Derivation of the position of the core of the air shower}

\begin{figure}
\centering
\includegraphics[width=\columnwidth]{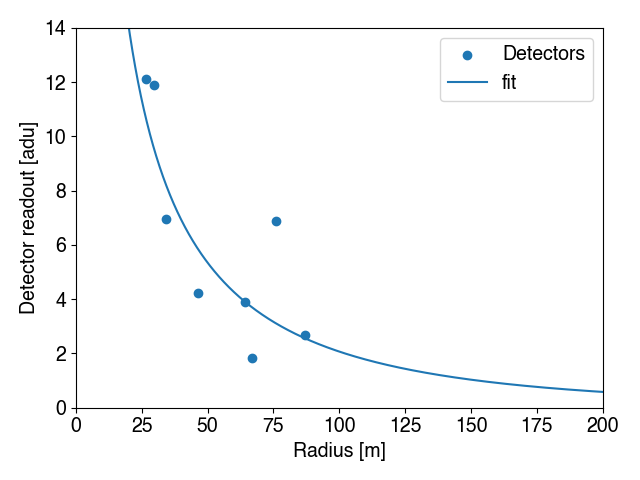}
\caption{Example of the LDF fit of Eq.~\ref{eq:ldf} to an example event with seven measured intensities. Uncertainties on the measurements are dominated by the Poisson statistics of detecting a given number of muons, with smaller contributions expected from the stochastic energy loss of individual muons, and contributions from the electromagnetic component, which we have not quantitatively estimated.}
\label{fig:Example_ldf}
\end{figure}

The altitude of 377.8\,m at the MWA site is relatively low, and so we expect the muonic component of EAS to be significant compared to the electromagnetic. We therefore use the approximation by \citet{1960ARNPS..10...63G} to the NKG lateral distribution function (LDF) of \citet{1958PThPS...6...93K} to model particle amplitudes as a function of perpendicular radial offset from the shower axis, $r$,
\begin{eqnarray}
\rho_{\rm NKG}(r,s,N_e)  =  \frac{N_e}{r_M^2} \frac{\Gamma(4.5-s)}{2 \pi \Gamma(4.5-2s)} \left( \frac{r}{r_M} \right)^{s-2} \left( 1+ \frac{r}{r_M} \right)^{s-4.5} \label{eq:ldf}
\end{eqnarray}
which --- while written in terms of the total electron number $N_e$ and Moliere radius $r_M = 420$\,m --- is expected to be applicable to the electromagnetic, muonic, and hadronic components for shower age parameters $0.5 < s < 1.5$. Given arrival direction fits from Section~\ref{sec:arrival_direction},we fit X,Y positions of the core (and hence radius $r$ for each detector), $N_e$, and $s$ from Eq.~\ref{eq:ldf} to the (negative) amplitudes of signals detected by the PDA, using our previously found conversion of 5\,ADC units per muon, and a total effective area of 0.81\,m$^2$. The fit only used detectors with a peak amplitude of at least -5\,ADC units (after compensating for DC offets), and minimised the sum squared differences between the logarithms of predicted and measured signal amplitudes. We note that off-zenith angles will decrease the effective area with $\sin$\,Alt, but the pathlength through the scintillator (and hence the signal amplitude) will increase as $(\sin$\,Alt$)^{-1}$, so we do not explicitly account for the Altitude in the fit (although it should be reflected in the mean shower age parameter, $s$). An example of the LDF fit is given in Figure~\ref{fig:Example_ldf}. For a reliable reconstruction, we required five detectors to contribute to a trigger, reducing the sample to 11,798 reconstructed events.

\begin{figure}
\centering
\includegraphics[width=\columnwidth]{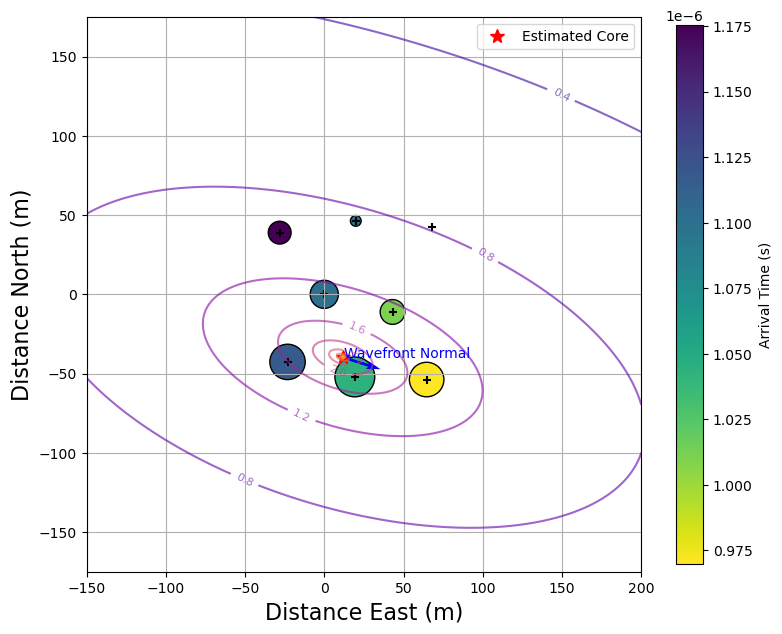}
\caption{Example of a reconstructed event. Detected signals are shown for each detector, with area proportional to amplitude, and color indicating relative arrival time in microseconds. The direction of the wavefront normal is given by the blue arrow, while the star indicated the estimated impact point. Purple lines correspond to regions of equal particle intensity.}
\label{fig:example_event}
\end{figure}

Figure~\ref{fig:example_event} gives an example of the event reconstruction, for a relatively well-reconstructed event for which seven of eight detectors registered a significant signal (ADC units > 5). The arrival times clearly indicate an arrival direction from the South--East, while the reconstructed core impact position is close to Detector 7.

\begin{figure}
\centering
\includegraphics[width=\columnwidth]{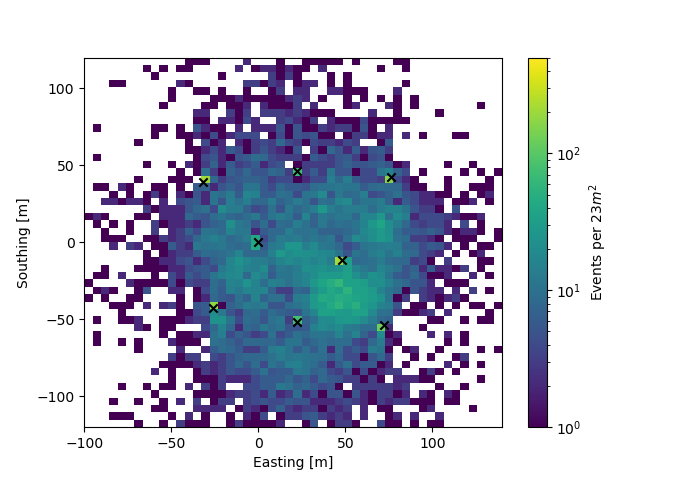}
\caption{Histogram of reconstructed impact positions on the array, showing the core region only.}
\label{fig:ground}
\end{figure}

Figure~\ref{fig:ground} gives the resulting reconstructed impact points for all events. As expected, the majority of detected events reconstruct close to the array --- either within, or a few tens of metres outside it. There are two obvious systematic effects: many events reconstruct with cores very close to each detector, and the South-East region, defined by detectors 5, 7, and 8, shows an excess of events. The first effect is due to the initial X,Y fit values being the position of the detector with the greatest amplitude, and the minimisation algorithm being stuck at that point. We have not been able to explain the second effect.

\subsection{Cosmic-ray energy spectrum}

\begin{figure}
\centering
\includegraphics[width=\columnwidth]{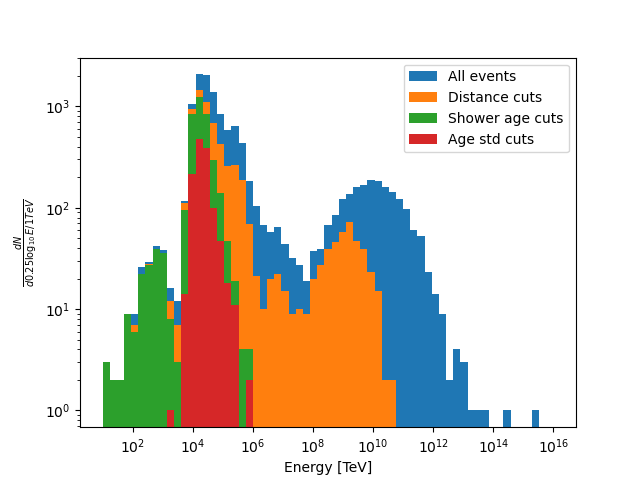}
\caption{Histogram of our energy reconstruction, showing successive cuts, from the initial sample of 11,798 events, after distance cuts (6147 events), cuts on shower age (3680 events), and cuts on relative shower age error (1272 events).}
\label{fig:energy_spectrum}
\end{figure}

We did not design our array to provide a precise energy reconstruction. Nonetheless, it is useful to derive an energy proxy, $E_\mu$, which may be useful for determining the trigger conditions on the MWA. We do this by integrating the fitted LDF (in units of muons) from a radius of 1\,m to 1500\,m, and ignore the electromagnetic contribution to get a total number of muons $N_\mu$. We then convert to total energy using \citet{2007BrJPh..37..633S}:
\begin{eqnarray}
\frac{E_\mu}{1\,{\rm TeV}} & = & \left(\frac{N_\mu}{9.8028}\right)^{1.172}.
\end{eqnarray}
The resulting distribution of our energy proxy $E_\mu$ is given in Figure~\ref{fig:energy_spectrum}. The peak of the distribution is close to 10\,PeV --- approximately what is expected, based on results from LORA \citep{LORA} --- but clearly there are events with spuriously high and low energies.

We note that our energy proxy $E_\mu$ likely over-estimates the true cosmic ray energy. While an EAS may be dominated by muons close to sea level, \citet{2016APh....85...50G} show how the ratio between muon number and electromagnetic energy, $r_{\mu e} = N(\mu)/(E_{\rm EM}/500\,{\rm GeV})$, never rises above $\sim 4$ even at very large shower depths \citep{2016APh....85...50G}. Furthermore, the fraction of electromagnetic energy deposited in a scintillator will be much higher, due to the greater energy loss rates of $e^{\pm}$, while the thin aluminium top of each detector does not provide much stopping power. We therefore attempt to estimate by how much $E_\mu$ over-estimates the true energy by obtaining a high-fidelity event sample, and comparing the measured to expected rates.

To obtain a high-fidelity sample, we place cuts as follows. Firstly, because our detector array is relatively small, we remove all events with cores reconstructed outside the array (i.e., outside a rectangle of area 103\,m EW by 90\,m NS), which removes those events reconstructed with very large distances and, hence, implausibly high energies. We next cut on the shower age parameter $s$, constraining it to physical values of $0.5 \le s \le 1.5$. This also removes very high-energy events, since a reconstruction with an artificially high $s$ will increase the implied energy for a given detection amplitude. Finally, we determine that well-fit cascades have a relative error on shower age of less than 30\% (i.e., $\sigma_s s^{-1} < 0.3$, which produces our final selection. This leaves 1272 events over two orders of magnitude in energy.

\begin{figure}
    \centering
    \includegraphics[width=\linewidth]{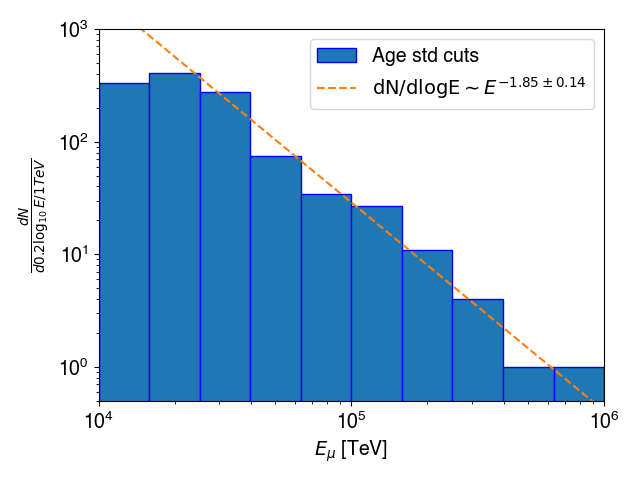}
    \caption{Spectrum of energy proxy assuming a purely muonic component, $E_\mu$, for EAS passing all selection cuts, compared to a power-law fit.}
    \label{fig:log_spec}
\end{figure}

\begin{figure}
    \centering
    \includegraphics[width=\linewidth]{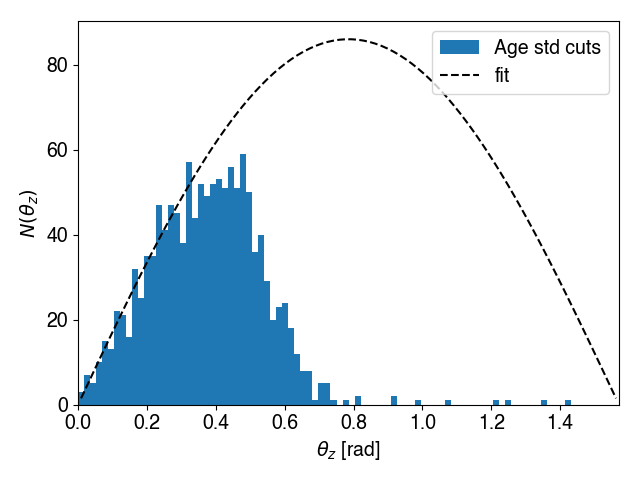}
    \caption{Histogram of zenith angle $\theta_z$ for events passing final selection cuts, compared to a fit of form $\sin \theta_z \cos \theta_z$.}
    \label{fig:final_thetaz}
\end{figure}

The estimated spectrum at each stage of these cuts is shown in Figure~\ref{fig:energy_spectrum}. The final sample has an estimated energy threshold of 6.3\,PeV, with only five events being reconstructed at lower energies. It peaks at 15.8\,PeV, and as shown in Figure~\ref{fig:log_spec}, has a power-law slope of $-2.85 \pm 0.14$ between 25\,PeV and 630\,PeV (beyond which we run out of statistics), consistent within 2$\sigma$ of the slope of the primary spectrum at these energies. Figure~\ref{fig:final_thetaz} shows the zenith angle distribution after these cuts --- the data follows the expected $\sin \theta_z \cos \theta_z$ dependence (due, respectively, to increasing solid angle, and decreasing projected area) over the range $0 \le \theta_z \le 0.35$, before atmospheric depth reduces the event rate. Thus we estimate that the particle detector array is fully efficient in the region $E_\mu > 25$\,PeV, $\theta_z < 0.35$, and the $103 \times 90$\,m$^2$ region of the core, in which we have 211 events. We do not attempt to model the efficiency outside of this range.

The nominal cosmic ray flux at $10^{16}$\,eV is approximately $4 \, 10^{-24}$\,m$^{-2}$\,s$^{-1}$\,sr$^{-1}$\,eV$^{-1}$ \citep[e.g.,][]{2024AdSpR..74.4403K}, with a power-law slope of $-3.1$. Within our completeness bounds, this implies a rate of only 4.2 particles per five days of observation. We interpret this difference as being due to our energy proxy $E_\mu$ being larger than the true energy by a factor of $\sim6.5$ (which would bring the observed and estimated rates into agreement), due to the contribution of the electromagnetic component to PDA signals. A secondary effect is that the relatively small size of our array forces us to fit our LDF close to the core, where fits are less reliable --- other detector arrays tend to characterise the LDF at larger values, e.g.\ above $\sim200$\,m for the Pierre Auger Observatory \citep{2005ICRC....7..291B}. We therefore use a final energy estimator of $\sim E = 0.15 E_\mu$, and conclude that our PDA is 100\% efficient in the parameter space $E > 4$\,PeV,  $\theta_z < 0.35$, and the $103 \times 90$\,m$^2$ core region.

\subsection{Implications for radio data}

The purpose of the MWA PDA is to detect particle cascades, and trigger radio data capture on the MWA. We have already verified the ability of the PDA to send artificially generated triggers to the MWA, and record eight-second segments of data containing the trigger time.\footnote{The eight-second segment duration corresponds to the size of individual files within the MWA buffer, and is thus the computationally simplest unit of data to work with. In the future, we expect to be able to return small portions of each file containing cosmic ray data.}

There is no dead-time associated with such a trigger, since the files containing radio information are first removed from the ingest loop of the MWAX correlator until the data is saved, at which point the files are freed to be returned to the loop. However, if too many files are simultaneously removed from the MWAX buffer, the smooth operation of the correlator could be compromised. We have not stress-tested this aspect of correlator functionality yet, but expect the limiting trigger rate to be of the order of once per minute; we currently aim for a trigger rate of approximately once per hour.

The raw trigger rate that produced our data-set averaged approximately 120 hr$^{-1}$, which is a factor of 100 higher than desired. However, given that MWA data stays in the buffer for over two minutes, the simple shower reconstruction performed here will be able to be run on each event, allowing a sub-selection to generate triggers. For instance, the rate for our post-cut sample of 1272 events is 4.3\,hr$^{-1}$, and 0.72 hr$^{-1}$ for our complete events.
Further tests with MWA data will be required to determine the optimal criteria for generating a trigger, and the acceptable data rate, though we expect pointing direction to be one of them.

\section{Conclusion}
\label{sec:conclusion}

We have designed, built, and deployed an array of eight particle detectors --- the MWA Particle Detector Array --- to the core of the MWA within Inyarrimanha Ilgari Bundara, the Murchison Radio-astronomy Observatory.

We have characterised the behaviour of both individual detectors, and the array as a whole, finding that the array can trigger on cosmic ray extensive air showers at a rate of approximately two per minute. A simplified reconstruction method, applied to a sample of 35,500 events recorded over a 12-day period, has (after quality cuts) produced a sub-sample of 1272 events with reliably reconstructed parameters, except for a systematic offset in estimated primary cosmic ray energy. This proves the ability of the MWA DPA to produce high-fidelity triggers to be generated and sent to the MWA to capture radio data from these events.

\paragraph{Acknowledgments}
This scientific work uses data obtained from Inyarrimanha Ilgari Bundara, the CSIRO Murchison Radio-astronomy Observatory. We acknowledge the Wajarri Yamaji People as the Traditional Owners and Native Title Holders of the observatory site. Support for the operation of the MWA is provided by the Australian Government (NCRIS), under a contract to Curtin University administered by Astronomy Australia Limited. We acknowledge the Pawsey Supercomputing Centre which is supported by the Western Australian and Australian Governments.

We would like to thank the KASCADE-Grande Collaboration for their donation of scintillator material and wavelength shifter for this experiment. We would also like to thank the Murchison Widefield Array Collaboration for many aspects of project support.

C.W.J.\ and J.E.D.\ acknowledge support from Australian Research Council Discovery Grant DP200102643. This project was funded through the Australian Research Council LIEF grant LE200100078.

\paragraph{Competing Interests}
The authors declare no competing interests.

\paragraph{Ethical Standards}
The research meets all ethical guidelines, including adherence to the legal requirements of the study country.

\printendnotes

\printbibliography

\end{document}